\documentclass[12pt]{article}

\usepackage{epsfig}

\title{A Search for Periodicity in the Super-Kamiokande
Solar Neutrino Flux Data}
\author{A.\,Milsztajn \\ DAPNIA-SPP, CEA Saclay \\ F-91191 Gif-sur-Yvette, 
France}

\begin{document} 
\setlength{\baselineskip}{4.0ex}

\maketitle

\begin{abstract}
The publicly available Super--Kamiokande 10 day plot data have been 
searched for a periodic modulation in the range between 10 and 500
days, with the periodogram method.  A peak is observed for 
$T = 13.75$~days, significant at 98.9~\% C.L.  I discuss possible
ways to check this result and, if confirmed, to investigate its
origin.
\end{abstract}

\vspace*{10mm}

%\newpage
%\section{Introduction} 

%\setlength{\baselineskip}{4.0ex}

\section{Introduction}

In recent years, we have witnessed a dramatic improvement in our 
assessment of the solar neutrino problem.  The very high statistics 
of solar neutrino interactions in ordinary water, obtained 
in the Super-Kamiokande experiment~\cite{sk1, sk2}, 
together with results from the chlorine~\cite{davis}, 
gallium~\cite{gallex, gno, sage} and heavy water~\cite{sno1,sno2}
detectors, have demonstrated that neutrinos 
undergo flavor conversion between the solar core and the earthly 
experiments.  
In parallel and at about the same time, the enormous progress in 
helioseismology and in solar models led to a firmer and refined
estimate of the emitted solar neutrino flux~\cite{tc2001},
that agrees perfectly with the measurements.
A few domains were singled out in the 
$\Delta m^2$ vs. $tan \theta$ parameter plane 
(see {\it e.g.} \cite{lisi01}), with 
the so-called Large Mixing Angle solution being the favored one.
The very recent observation of reactor neutrino oscillations by the
KamLAND group~\cite{kaml} has further pinned down this solution, 
although the allowed domain in parameter space is now comprised of 
two disjoint sub-domains (see {\it e.g.} \cite{lisid02}).  

In the present standard view, no large periodic modulation 
is expected in the solar neutrino signal.  Two small effects 
might be observed, with periods of one year and one day.
The obvious yearly modulation is due to the Earth's 
orbit excentricity, $e = 0.0167$, which induces a flux variation 
$ 4 e / (1-e)^2 \simeq 7 \% $ between aphelion and perihelion.  
There is now some evidence for this flux modulation, at a modest 
significance level~\cite{sk1}.  The possible 
daily periodicity stems from neutrino oscillations inside 
the Earth~\cite{MSWearth}. The amplitude of this so-called day/night 
effect depends strongly on the neutrino masses and mixings. 
The day/night comparisons by the Super-Kamiokande~\cite{sk1,sk2} and 
SNO~\cite{sno2} collaborations helped to reject some 
solutions to the solar neutrino problem, and to narrow down the 
allowed parameter space domain. 
% It is still worth looking for
% it, as the expected day/night effect is different in the two
% sub-domains of the LMA solution. 

Searches for other periodicities have also been performed, such as
correlations with the 11-year solar cycle ({\it e.g.} \cite{cycle}), 
or the solar synodic period (see {\it e.g.} \cite{stur1, stur2} 
and references therein); it seems fair to say that no 
consensus exists on the presence of such periodic modulations.  

Very recently, in December 2002, the Super-Kamiokande collaboration 
has released its solar neutrino flux data in 10 and 45 day bins 
over the period May 1996 to July 2001, 
corresponding to 22,400 solar neutrino interactions (see
{\tt http://www-sk.icrr.u-tokyo.ac.jp/sk/lowe/index.html}).
This makes it possible to perform a systematic search for
any periodicity in this data set, not restricted to the 
four periods listed above. 
The present paper reports the results of such
a search in the period range from 10 to 500 days, as well as
evidence for a periodic modulation, at $T = 13.755 \pm 0.011 $~days. 
I am not aware of any previous systematic periodicity search on
solar neutrino fluxes in a wide period range, and especially 
below 18~days\footnote{Half a day after this preprint was first
posted, I became aware of a similar search by P.A. Sturrock 
and D.O. Caldwell, through an AAS meeting abstract, posted at 
{\tt http://www.aas.org/publications/baas/v34n4/aas201/647.htm}~.
I very much thank T.~Lasserre for pointing this out to me. 
The periodicity found in the present paper agrees perfectly
with that given in this conference abstract.}

\section{The data sample and the search for periodicity}

I have used the Super-Kamiokande solar neutrino flux in 10 day bins only, 
in order to probe the shorter periods.  
There are $N = 184$ such data points, for each of which are provided 
the date and time of the beginning and end of the 10-day bin (in Japan
Standard Time JST), the value of $d^2$, the average squared 
Sun-Earth distance in this bin, and the boron-8 solar neutrino flux, 
together with errors, uncorrected for this distance variation.  
The relative flux errors are typically between 10 and 20\%.

The total time span of the data is 1,864 days (184 bins of 10.1~days
in average), corresponding to an actual live time of 1,496 days. 
The bin size is rather constant, as only 13 bins are shorter than
8 or longer than 12~days, and the standard deviation of the bin size
distribution is 1.27~day.  The actual distribution of dead time
is not provided, so that I have chosen to assign the neutrino flux 
value to the center of each bin.  I have then corrected this flux value
for the Sun-Earth distance, using in effect a flux rescaled at 1 AU.
(I have used the $d^2$ factor provided in the data table, and have checked
that it agrees with that computed using the JPL DE406 Solar System 
ephemerides available at {\tt http://ssd.jpl.nasa.gov/}~.) Finally,
I subtract in each bin the average flux value given in the same table.

I have then performed a search for periodicity on these ``flux excursions'', 
using a slight variant of the periodogram method~\cite{lomb,scargle}.  
All flux values were used with equal weights, {\it i.e.} ignoring
their errors.  The frequency range from 
0.002 to 0.100~d$^{-1}$ was searched, with a frequency step of 
0.0001~d$^{-1}$; this corresponds to a period range between 500 
and 10~days.  (The most significant peaks were then scanned again, 
with a 10 times smaller step, 0.00001~d$^{-1}$.)
The resulting power spectrum is shown in Fig.~1, where the 
plotted variable is 
$$ z(\nu) \, = \, \frac{ | \, \sum_i \, (\Phi_i - \Phi_{av}) \, 
      exp \, ( 2 \; i \; \pi \; \nu \; t_i ) \,
      |^2}{ \sum_i \, (\Phi_i - \Phi_{av})^2 } \, , $$
where $\Phi_i$ is the flux measured at time $t_i$, $\Phi_{av}$ the
average flux and $\nu$ is the probed frequency; the sums run over
the $N$ data points.
%
%    --------------     FIGURE 1      ------------------------------
%
% \begin{figure} [ht] 
%  \begin{center} \epsfig{file=fig1_sk10.eps,width=11.8cm}
% % 7.8 cm for article 
% % \vspace{-.1cm}
%   \caption{The power spectrum in the Super-Kamiokande 10 day plot data,
%    from a periodogram analysis.  The levels corresponding to a
%    1, 5 and 10 percent probability of a statistical fluctuation
%    are indicated.   
%    }  
%   \label{periodo}
%  \end{center} 
% %\vspace{-0.7cm}
% \end{figure}
%
%    --------------     FIGURE 1     ------------------------------
%
%
%    --------------     FIGURE 2      ------------------------------
%
% \begin{figure} [ht] 
%  \begin{center} \epsfig{file=fig2_sk10.eps,width=11.8cm}
% % 7.8 cm for article 
% % \vspace{-.1cm}
%   \caption{The $z$ distribution corresponding to Fig.~1. For data with
%    no significant periodicity, this distribution should be
%    exponential ($ \propto e^{-z} $). The three z values in excess of
%    6.5 belong to the same periodogram peak at 
%    $ \nu = 0.07270 \mathrm{d}^{-1} $.
%    }  
%   \label{expo}
%  \end{center} 
% %\vspace{-0.7cm}
% \end{figure}
%
%    --------------     FIGURE 2     ------------------------------
%

Only one prominent peak is found, at a frequency 
$ \nu = 0.07270 \, \mathrm{d}^{-1} \simeq 842 $~nHz, 
corresponding to  $ T = 13.75 \pm 0.05 $~days. 
(This error is only an estimate, from the width 
of the periodogram peak.)  The height of the peak is $ z_0 = 10.4 $,
corresponding to a probability $p = e^{-z_0} \simeq 3 \cdot 10^{-5}$.
The $z$ distribution is shown in Fig.~2.
In order to estimate the probability that such a peak be compatible
with a statistical fluctuation, one has to know the number of
independent frequencies tested by the periodogram analysis. For the
frequency range tested here, essentially between 0 and twice the
Nyquist frequency, this number is close to $ 2 N $, {\it i.e.} twice 
the number of data points (see {\it e.g.} \cite{numrec}, section
13.8, and references therein).  The probability to obtain a peak 
height greater than $z_0$ from $ 2 N $ tested frequencies is 
$ 1 - ( 1 - p)^{2N} \simeq 2 N p \simeq 1.1 \% $.
Thus the observed peak corresponds, at $98.9 \%$~C.L., to a 
periodicity in the Super-Kamiokande 10 day plot data (as of December
2002).  The second highest peak of the periodogram, with 
$z_1 = 6.4 \, $, is perfectly compatible with statistical noise at 
$47 \%$~C.L.

Note that the sampling, though quite regular, is sufficiently 
variable that no aliasing is observed in the periodogram.
%
%    --------------     FIGURE 3      ------------------------------
%
% \begin{figure} [ht] 
%  \begin{center} \epsfig{file=fig2_sk10.eps,width=11.8cm}
% % 7.8 cm for article 
% % \vspace{-.1cm}
%   \caption{The flux values versus their phase in a period $T = 13.75$~d.
%    The phase is here the fraction of a period T, counted from an arbitrary
%    origin (MJD 0 $\equiv$ JD 2,450,000). 
%    The phase is a number between 0 and 1, but
%    each data bin has been plotted here twice, at its phase and phase~+~1.
%    }  
%   \label{lc}
%  \end{center} 
% %\vspace{-0.7cm}
% \end{figure}
%
%    --------------     FIGURE 3     ------------------------------
%

The folded flux values are plotted versus their phase in Fig.~3.
The phase here is defined as the fraction of a period $T$ counted 
from an arbitrary origin (Modified Julian Day 0, or JD 2,450,000
$\equiv$ 1995, Oct 9.50~UTC); it is a number between 0 and 1.  
The modulation is clearly visible on Fig.~3.

In order to better estimate the parameters and their errors,
I have fitted the 184 data points $ ( \Phi_i, \sigma_{\Phi_i} ) $
with a sinusoidal flux modulation, 
$$ \Phi_0 \, \, [ \, 1 \, + \, A \, sin ( 2 \pi \, \frac{t_i - 1200}{T} 
  \, + \, \phi_{1200} ) \, ] \, ,$$
where $ \Phi_0 $ is the time-averaged $^8$B neutrino flux, 
$A$ the relative
signal modulation, $T$ the period and $\phi_{1200}$ the phase at a
reference date, MJD 1200.0, chosen close to the median time
of the SK data.
In this fit, the points were weighted using the inverse squared errors
quoted by the Super-Kamiokande collaboration.
The resulting $\chi^2$ is 165.8 for $(184-4)$ degrees of freedom.
The fitted parameters values are~:
$\Phi_0 = 2.329 \pm 0.024 \, 10^6 \; ^8{\mathrm B-} \nu \, 
cm^{-2} \, s^{-1}$,
$ A = 0.067 \pm 0.014 $, $ T = 13.755 \pm 0.011 $~days and 
$\phi_{1200} = 0.55 \pm 0.22$~rad.  (In comparison,
the $\chi^2$ of a constant fit is 187.7 for 183 d.o.f.)

It may seem strange to be able to detect a period that is only
35~\% longer than the average time bin size. The neutrino flux
in each bin is integrated over the bin width, so that any periodic
signal will be washed out when integrated over $ \simeq 3/4$ of a 
period.  In the case of a purely sinusoidal modulation, with 
amplitude $A$ relative to the average flux, the measured 
amplitude after signal averaging over a random 3/4 fraction of a 
period is $ A' = (2^{3/2} / 3 \pi) A \simeq 0.30 A $.  I will
discuss this further in the next section.  

Note that, for such an integrated signal, the periodogram search 
has a much reduced sensitivity to periods close to the time
sampling, 10.1~days, and twice the time sampling, 20.2~days.
Thus the number of independent frequencies, $2 N$, that was used 
above is probably a bit overestimated, and the signal significance 
under-estimated. A full Monte-Carlo simulation would be necessary
to quantify this. 

In order to search for a possible bias, either on the 13.75 day
period or on its significance level, I have performed a Monte-Carlo
simulation generating 10,000 experiments with exactly the same 10 day 
time bins as Super-Kamiokande, and flux values taken at random 
from a normal law with the same average 
($2.35 \cdot 10^6 \, \nu \, \mathrm{cm}^{-2} \, \mathrm{s}^{-1} $)
and the same dispersion ($0.33 \cdot 10^6 $) 
as the 184 time bins.  The observed $z$ distribution is in perfect 
agreement with an exponential law, and no significant bias is 
observed near 13.75~days, or anywhere else in the periodogram.
Out of the 10,000 simulated data sets, 123 have a maximum $z$ value 
larger than $z_0 = 10.4 \;$. This agrees with our confidence
level estimate of 98.9\%.

I have also performed a periodogram search on the combined
results of the {\sc gallex}~\cite{gallex} and {\sc gno}~\cite{gno} 
collaborations over the period 1991-2001.  
No significant peak was found.
This gives no information on a possible 13.75~day periodicity
however~: out of the 100 {\sc gallex} and {\sc gno} runs, 65 
correspond to an exposure time very close to 27.5~days, {\it i.e}
twice 13.75~days, so that a much reduced sensitivity is expected.
I have not analysed the chlorine data, as the time between 
extractions is even longer for this experiment.

\section{Discussion and Future Work}

Provided the signal presented in the previous section is not
simply due to a statistical fluctuation, there exist many tests
that could be performed to verify its existence.  An identical
analysis using the same Super-Kamiokande data, but binned in
3-4~days bins would already prove very useful. The statistics
in each bin would decrease by a factor 3, and the statistical 
errors would increase typically by $\sqrt{3}$. The distribution
of measured fluxes would hence be wider.  But a 13.75~day
periodic signal would then, in these bins, be integrated over
1/4 of a period instead of 3/4. The amplitude of the signal 
would increase by a factor 3, to about 20\%,
thus more than compensating the
increase in the width of the distribution; the significance of
any real modulation could be clearly established. 

It would then be interesting to check what the actual period is, 
as 13.75~days could very well be the (dominant) second harmonic
of a 27.5~day period. (The periodogram method is not well suited
in such a case; the AoV method \cite{alex} could be used instead.)
One should also check for aliases with periods shorter than
10~days.

A search for periodicity is always more efficient starting from
unbinned data.  The ultimate verification would be to search for
periodicity using information on each ``candidate solar
neutrino'', {\it i.e.} the time of detection, the angle with 
respect to the Sun ($cos \theta_{\odot}$) and the energy.
For the Super-Kamiokande data, one can only speak of 
``candidate solar neutrinos'' as the boron-8 solar neutrino 
flux is measured as an excess over the roughly constant
background in the $cos \theta_{\odot}$ distribution (see for example 
figure 1 of \cite{sk3}).  One could use
a modified version of a periodogram analysis, where each ``candidate''
would receive a $cos \theta_{\odot}$-dependent weight, equal to the
fraction of solar neutrinos at its measured $cos \theta_{\odot}$.
(The closer $\theta_{\odot}$ to zero, the larger the weight.)

A simplified Monte-Carlo simulation of such a weighted periodogram 
analysis has been performed~\cite{me}.  
It shows that a search for an unknown
periodic modulation of the flux in the period range from a few
hours to a year would allow to detect a 10~\% modulation at
better than 99~\%~C.L. (The confidence level increases very fast
with the amplitude.)  This means that, for a fixed confidence level,
the searched frequency interval would then be about a hundred times 
greater than in the present paper. Equivalently, examining in such 
an analysis only periods greater than 10~days would lead to confidence
levels 100~times better. 

In case the signal is confirmed, this would not tell us whether it is
due to an instrumental effect, to an actual modulation of the boron-8 
neutrino flux, or if it is induced by a modulation in the background. 
The latter hypothesis could be tested, 
for example, by a periodogram analysis of events that
enter the $cos \theta_{\odot}$ plot at negative values (the anti-solar
hemisphere)~: they are almost all background events. It would also be 
interesting to search for the same periodicity in other Super-Kamiokande
data sets, such as atmospheric neutrinos.

One cannot help noticing that 13.75~days is very close to half the
solar synodic rotation period. This may turn out to be a coincidence,
or not. There exist many solar observables that display a 13-14~day
periodicity, although the peak in their Fourier spectrum is
usually quite wide.  Examples of these are properties of the chromosphere, 
solar wind characteristics at the Earth, and even induced geomagnetic
activity (see {\it e.g.} \cite{mursula} and references therein). 
(The latter two might influence
the background level in underground experiments; a quick bibliographic 
search did not allow me to find a reference in the litterature to a 
periodicity in the flux of cosmic rays.)  Note however that the
observed 13.5~day periodicities have not been seen to persist
over a very long time span, such as the 5 years 
of data analysed here.
Moreover, the intensity of these modulations vary during the 11-year
solar cycle, and often show phase changes between consecutive periods
of 13.5~day variability.  For that reason, I have tried to split
the Super-Kamiokande data into various time bins to search for 
such phase shifts, with no positive results.  It is likely however
that the 10-day binned data sensitivity is not large enough to
detect such effects, and that the periodogram is not the proper tool 
here.

In case there is a connection between the observed signal and solar
properties, it would certainly be interesting to search for a 
correlation with the position of the Earth relative to the Sun's
heliographic or heliomagnetic latitude.

Regarding other experiments, we
have already mentioned that the gallium and chlorine data
have a lower sensitivity, due to the long time between extractions,
and to an unfortunate coincidence. If the Super-Kamiokande signal is
confirmed by further analyses, the matter should of course be 
reconsidered, as one would then just have to study the  
amplitude of a modulation at a known period. In contrast,
all other solar neutrino experiments with ``real time detection'' 
such as SNO, or the forthcoming Borexino~\cite{borexino}, 
can use the techniques described above or similar ones
to search for a signal.  One would expect the SNO data to display 
the same kind of variation as the SK data, at the level of 20\%.

\section{Summary}

I have presented the results of a search for a periodic modulation
in the Super-Kamiokande 10 day plot data (as of December 2002).
At 98.9~\% C.L., a $ T = 13.75 \pm 0.01 $~day modulation is found.
This is close to half the Sun's synodic rotation period.   
I have discussed methods to soon check this result, and have 
sketched possible future directions, in case it is confirmed.

\vspace{12mm}
{\bf Acknowledgments}

I am grateful to the Super-Kamiokande collaboration for the release 
of the 10 day plot data, and especially to Dr. Nakahata for kindly 
answering my questions about it.
I am indebted to Michel Cribier for numerous discussions about solar
neutrinos, and for his help with the {\sc gallex} and {\sc gno} data.
I thank Jacques Bouchez, Thierry Lasserre, Bruno Mansouli\'e 
and Jim Rich for discussions and comments on this manuscript.  
I thank Patrick Tisserand for directing
me to important references on the known periodicities of the Sun in the
13-14~days range, and for his help in preparing this paper.

\newpage*
 
\begin{center}
Figure Captions
\end{center}
 
\vspace*{5mm}
 
\noindent
1. The power spectrum in the Super-Kamiokande 10 day plot data,
   from a periodogram analysis.  The levels corresponding to a
   1, 5 and 20 percent probability of a statistical fluctuation
   correspond here to $z = 10.5$, 8.9 and 7.4, respectively.   
 
\vspace*{15mm}
 
2. The $z$ distribution corresponding to Fig.~1. For data with 
   no significant periodicity, this distribution should be
   exponential ($ \propto e^{-z} $). The three $z$ values in 
   excess of 6.5 all belong to the same periodogram peak at 
   $ \nu = 0.07270 \, \mathrm{d}^{-1} $.

\vspace*{15mm}
 
3. The flux values versus their phase in a period $T = 13.75$~d.
   The phase here is the fraction of the period $T$, counted from an 
   arbitrary origin (MJD 0.0 $\equiv$ JD 2,450,000). 
   It is a number between 0 and 1, but each data bin has been 
   plotted twice, at its phase and phase~+~1.

\newpage*
 
\begin{figure}[h]
\centering\psfig{file=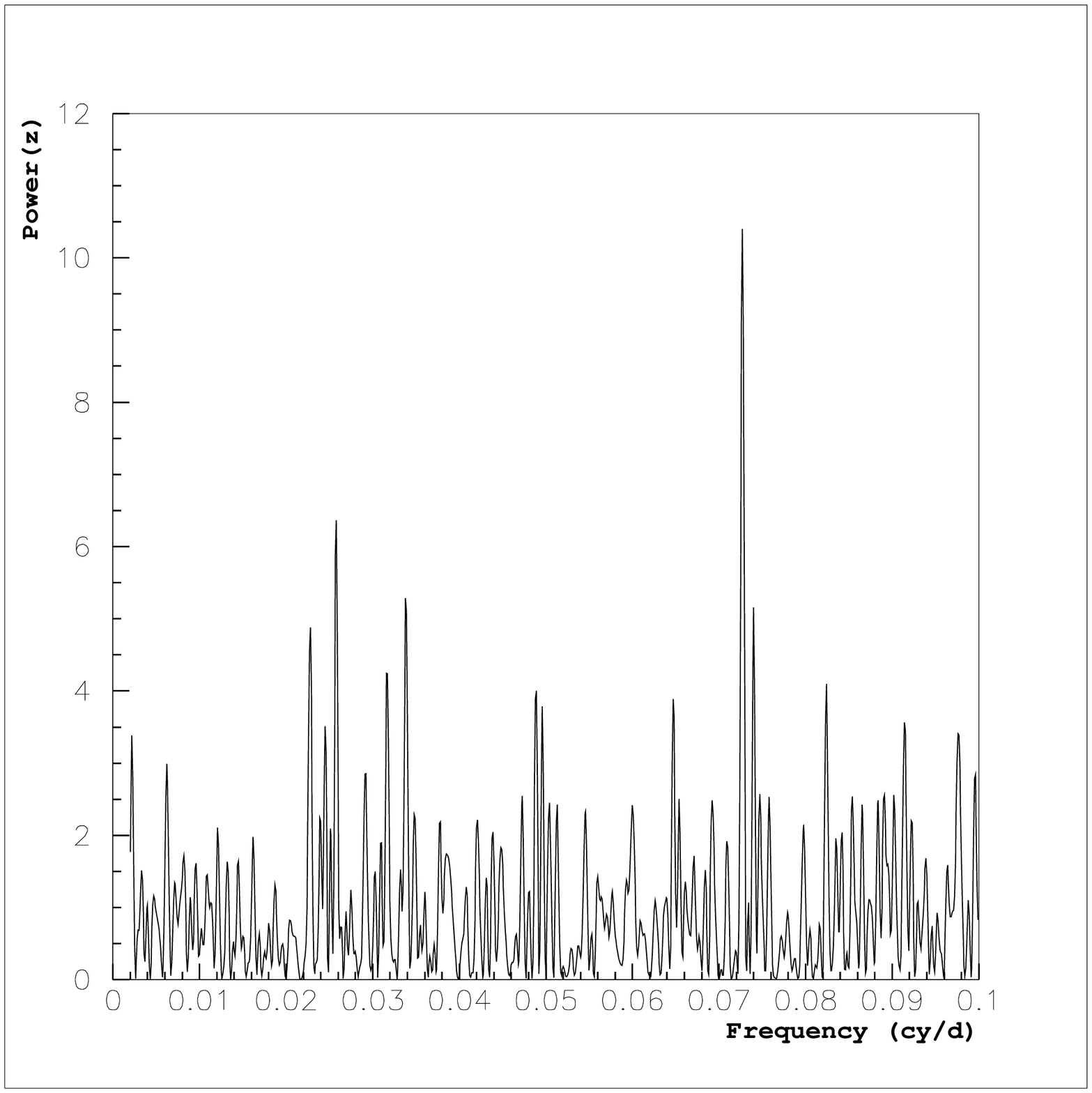,width=\textwidth}
\label{periodo}
\end{figure}

\newpage*
 
\begin{figure}[h] 
\centering\psfig{file=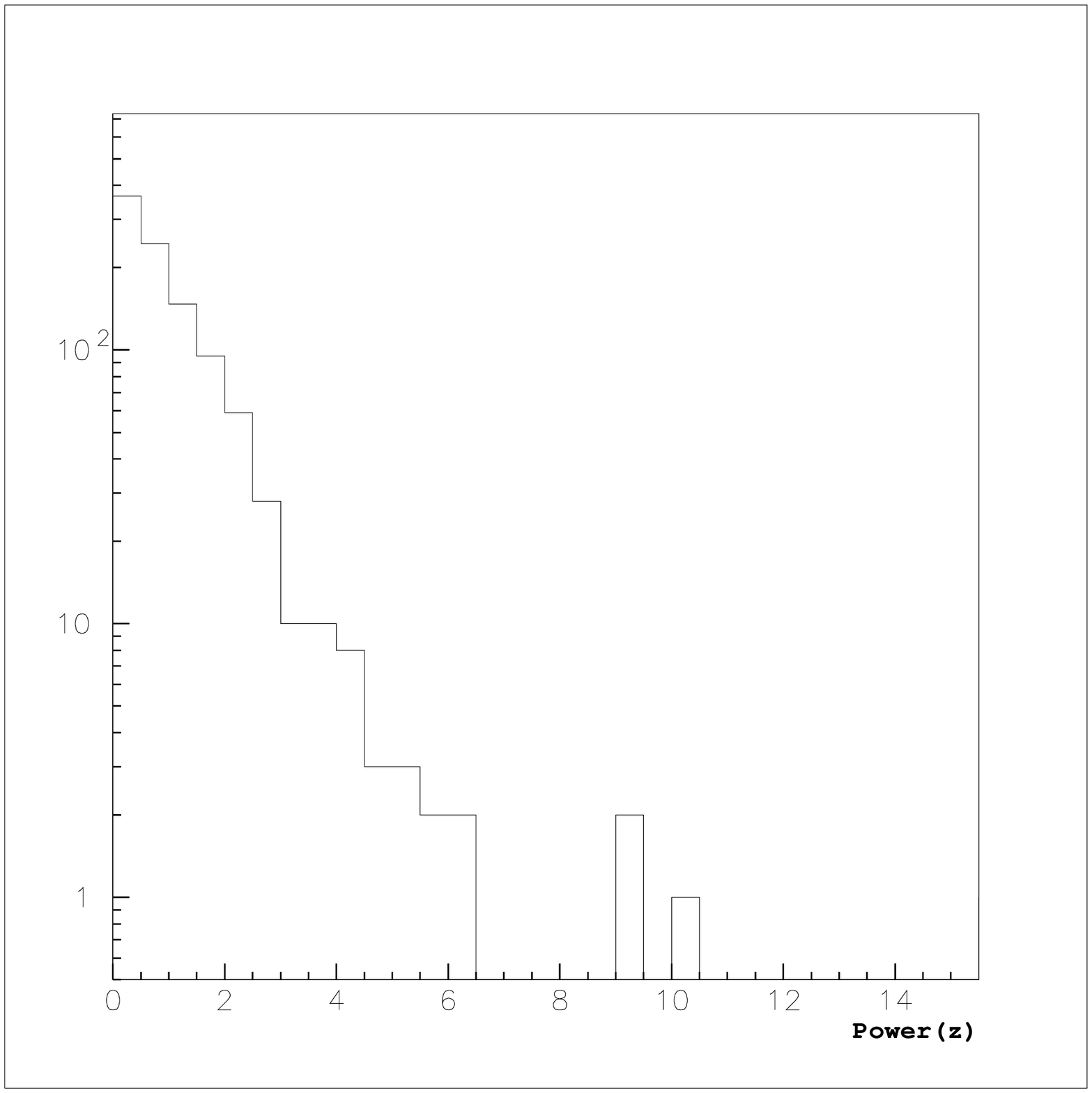,width=\textwidth}
\label{expo}
\end{figure}
 
\newpage*
 
\begin{figure}[h] 
\centering\psfig{file=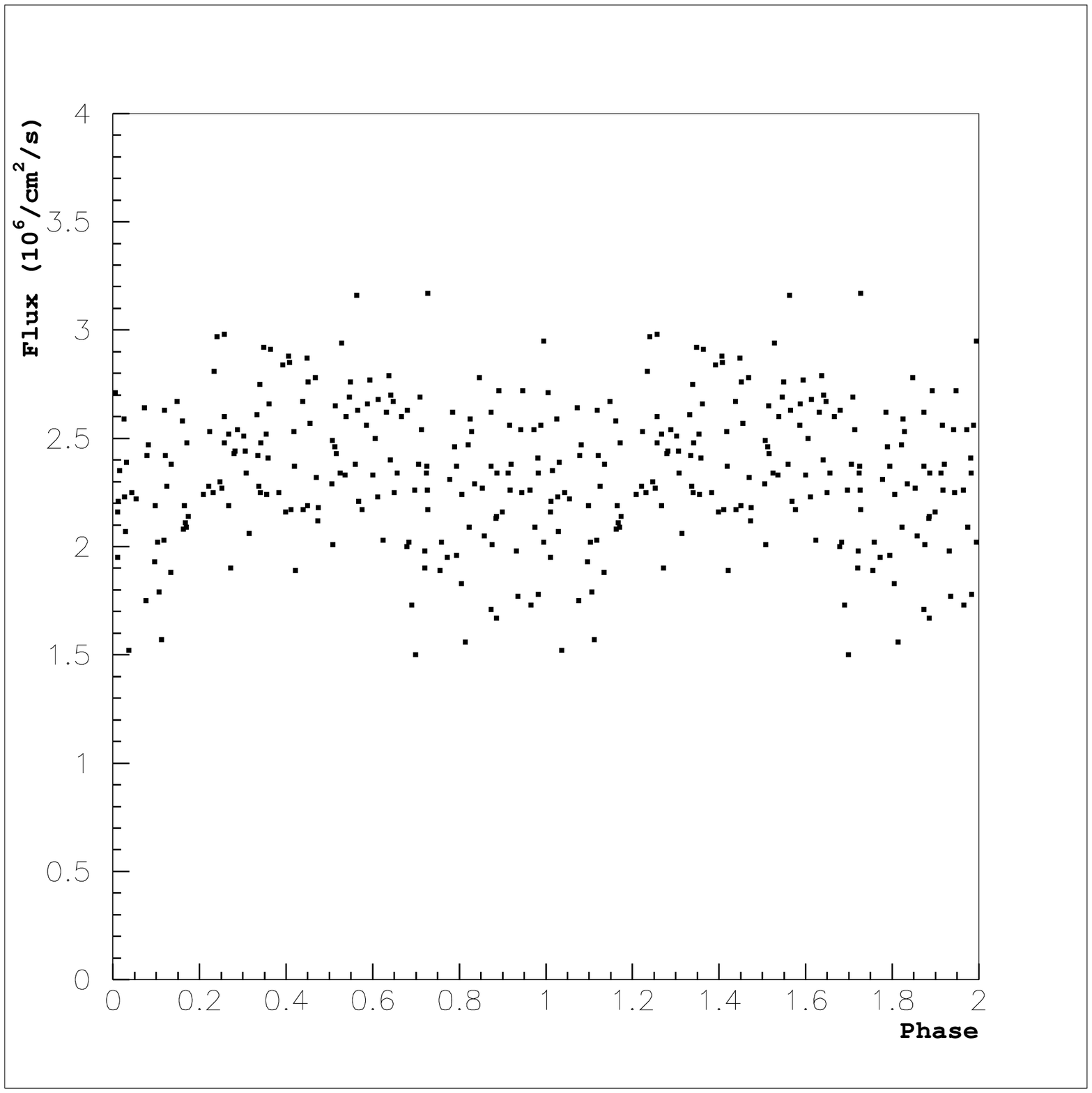,width=\textwidth}
\label{lc}
\end{figure}


\begin{thebibliography}{99}  
\itemsep 0pt
\parsep 0pt   
\bibitem{sk1} S.~Fukuda et al., Phys. Rev. Lett. {\bf 86} (2001) 5651. 

\bibitem{sk2} S.~Fukuda et al., Phys. Lett. {\bf B 539} (2002) 179. 

\bibitem{davis} B.T.~Cleveland et al., Astrophys. J. {\bf 496} (1998) 505. 

\bibitem{gallex} W.~Hampel et al., Phys. Lett. {\bf B 447} (1999) 127. 

\bibitem{gno} M.~Altmann et al., Phys. Lett. {\bf B 490} (2000) 16.

\bibitem{sage} J.N.~Abdurashitov et al., J. Exp. Theor. Phys. {\bf 95} 
(2002) 181. 

\bibitem{sno1} Q.R.~Ahmad et al., Phys. Rev. Lett. {\bf 89} (2002) 011301  

\bibitem{sno2} Q.R.~Ahmad et al., Phys. Rev. Lett. {\bf 89} (2002) 011302  
 
%\bibitem{sno2} Q.R.~Ahmad et al., Phys. Rev. Lett. {\bf 87} (2001) 071301 ?? 

\bibitem{tc2001} S.~Turck-Chi\`eze et al., Astrophys. J. {\bf 555} 
(2001) L69. 

\bibitem{lisi01} G.L.~Fogli et al., Phys. Rev. {\bf D 66} (2002) 093008 

\bibitem{kaml} K.~Eguchi et al., Phys. Rev. Lett. {\bf 90} (2003) 021802 

\bibitem{lisid02} G.L.~Fogli et al., preprint astro-ph/0212127 

\bibitem{MSWearth} J. Bouchez et al., Z. Phys. {\bf C32} (1986) 499. 

\bibitem{cycle} J.N.~Bahcall, G.B.~Field and W.H.~Press, Astrophys. J. 
{\bf 320} (1987) L69. 

\bibitem{stur1} P.A.~Sturrock and M.A.~Weber, Astrophys. J. {\bf 565} 
(2002) 1366. 

\bibitem{stur2} P.A.~Sturrock and J.D.~Scargle, Astrophys. J. {\bf 550} 
(2001) L101. 

\bibitem{lomb} N.R.~Lomb, Astrophys. and Space Sci. {\bf 39} (1976) 447. 

\bibitem{scargle} J.D.~Scargle, Astrophys. J. {\bf 163} (1982) 835. 

\bibitem{numrec} W.H.~Press et al., Numerical Recipes in C~: The Art 
of Scientific Computing, Cambridge University Press (1992). 

\bibitem{alex} A.~Schwarzenberg-Czerny, Month. Not. Roy. Astr. Soc.
 {\bf 241} (1989) 153. 

\bibitem{sk3} M.~Smy (Super-Kamiokande coll.), preprint hep-ex/0208004.

\bibitem{me} A.~Milsztajn, manuscript in preparation.

\bibitem{borexino} G.~Alimonti et al., Astropart. Phys. {\bf 16} (2002) 205. 

\bibitem{mursula} K.~Mursula and B.~Zieger, Journal of Geophys. Res. 
{\bf 101}, A12 (1996) pp 27077-27090. 


\end{thebibliography}
\end{document}